# ARMAN: A Reconfigurable Monolithic 3D Accelerator Architecture for Convolutional Neural Networks

Ali Sedaghatgoo, Amir M. Hajisadeghi, Mahmoud Momtazpour, and Nader Bagherzadeh

**Abstract**—The Convolutional Neural Network (CNN) has emerged as a powerful and versatile tool for artificial intelligence (AI) applications. Conventional computing architectures face challenges in meeting the demanding processing requirements of compute-intensive CNN applications, as they suffer from limited throughput and low utilization. To this end, specialized accelerators have been developed to speed up CNN computations. However, as we demonstrate in this paper via extensive design space exploration, different neural network models have different characteristics, which calls for different accelerator architectures and configurations to match their computing demand. We show that a one-size-fits-all fixed architecture does not guarantee optimal power/energy/performance trade-off. To overcome this challenge, this paper proposes ARMAN, a novel reconfigurable systolic-array-based accelerator architecture based on Monolithic 3D (M3D) technology for CNN inference. The proposed accelerator offers the flexibility to reconfigure among different scale-up or scale-out arrangements depending on the neural network structure, providing the optimal trade-off across power, energy, and performance for various neural network models. We demonstrate the effectiveness of our approach through evaluations of multiple benchmarks. The results demonstrate that the proposed accelerator exhibits up to 2x, 2.24x, 1.48x, and 2x improvements in terms of execution cycles, power, energy, and EDP respectively, over the non-configurable architecture.

——————————— ◆ ———————————

## 1 INTRODUCTION

The Convolutional Neural Networks (CNNs) have gained widespread popularity and adoption, revolutionizing the field of deep learning. Their automatic feature detection and extraction capability without human supervision has been instrumental in a variety of applications such as face detection, image classification, search, and machine vision research. A CNN consists of two primary processes: training and inference. The training process is performed once to learn meaningful features from the data, while the inference process applies the acquired knowledge repeatedly. The inference process requires extensive multiplication and addition operations, leading to high energy consumption and latency [1]. Furthermore, this huge amount of Multiply and Accumulate (MAC) operations necessitates massive off-chip memory accesses, which increases overall energy consumption and processing delay [2].

Currently, Graphics Processing Units (GPUs) are commonly employed for the training and inference of neural networks. However, GPUs exhibit limitations such as high power consumption, considerable area requirements, and sluggish inference performance, making them unsuitable in many applications such as edge computing and the Internet of Things (IoT) [3]. In this regard, hardware accelerators have been developed, enabling reduced energy consumption, and high computation speed for CNN inference. Nevertheless, hardware accelerators face some other challenges, including incompatibility with different algorithms, lack of scalability, and low reconfigurability to effectively support different algorithms and NN models [4].

In light of these challenges, improving the performance and utilization of CNN accelerators become crucial. However, as we demonstrate in this paper via extensive design space exploration, different neural network models have different characteristics, which calls for different accelerator architectures and configurations to match their computing demand. For example, a neural network model with a specific number and size of layers may perform very well on a specific scale-out (e.g., 2x2) systolic-array-based accelerator, but perform poorly on another one. To overcome this challenge, this paper presents a reconfigurable architecture based on systolic arrays to accelerate the inference of CNNs by leveraging M3D technology. The proposed architecture ensures performance, power, energy, and EDP improvements when necessary for different neural network models by offering reconfigurability of different scale-out arrangements. Moreover, realizing this architecture on the M3D platform enables low communication overhead within the reconfigurable architecture, resolving issues such as long inter-component communications and large areas.

The main contributions presented in this paper are threefold:
- We provide a Pareto-based design space exploration framework to explore the multi-objective design space of the problem.

———————————————
*Ali Sedaghatgoo, Amir M. Hajisadeghi, and Mahmoud Momtazpour are with the Department of Computer Engineering, Amirkabir University of Technology (Tehran Polytechnic), (e-mail: {alisedaghatgoo, hajisadeghi, momtazpour}@aut.ac.ir).*
*Nader Bagherzadeh is with the Department of Electrical Engineering and Computer Science, University of California Irvine, (e-mail: nader@uci.edu).*

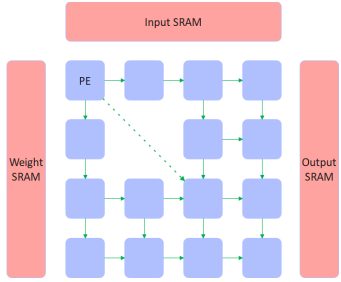

Fig. 1. A systolic array structure.

- We propose a reconfigurable systolic array-based CNN accelerator architecture based on the explored design space, that can effectively rearrange the systolic array in run-time to reach optimal execution time, energy, power, or EDP when necessary.
- We provide an extensive evaluation to verify the effectiveness of the proposed architecture to reach optimal performance, power, energy, or EDP for accelerating CNN inference.

This paper proceeds with a brief introduction to the key concepts employed in our research in section 2. Subsequently, section 3 describes our motivation for conducting this research, followed by the multi-objective problem, design space exploration, and the proposed accelerator architecture. The experimentation process and results are presented in section 4, followed by a discussion of previous research in section 5. Finally, the conclusion drawn from this study is presented in section 6.

## 2 BACKGROUND

### 2.1 Convolutional Neural Network

CNNs have revolutionized computer vision, but their increasing complexity and computational demands pose challenges for real-time applications. Certain layers in CNNs, such as convolutions and fully connected layers, contribute significantly to the computational burden [5]. Efficient computation mapping techniques aim to optimize these layers by mapping them onto specialized hardware platforms, maximizing parallelism and resource utilization [6]. This involves minimizing execution time through parallelized and pipelined matrix multiplications and additional operations. By focusing on compute-intensive layers and developing efficient mapping strategies, CNN accelerators can greatly enhance performance and utilization.

### 2.2 Systolic Array

Systolic arrays are specially-designed hardware structures that efficiently perform parallel and pipelined computational tasks. As shown in Fig. 1, they consist of processing units, SRAMs, and communication networks. These arrays allow for parallel movement of data through input and output processing units, resulting in rapid processing and improved operational utilization. Systolic arrays are capable of executing specific operations on incoming data and transmitting the results to the output processing units. Their flexibility and potency make them suitable for a wide range of applications, including signal and image processing, numerical matrix computations, encryption techniques, data compression, and machine learning [7]. With their ability to handle parallel and pipelined MAC operations effectively, systolic arrays are ideal choices for neural network inference [8].

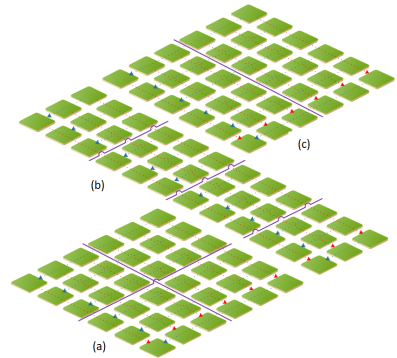

Fig. 2. Systolic array arrangements: a) 2×2, b) 4×1, and c) 2×1H.

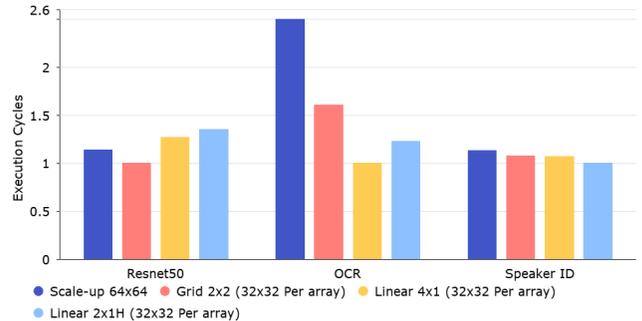

Fig. 3. Execution cycle required for each of the benchmarks.

### 2.3 Monolithic 3D

The M3D technology is an important advancement in chip manufacturing. It enables the stacking of multiple chip layers on top of each other and connecting them vertically [9]. This is different from the traditional TSV-based technology that connects multiple wafer chips. On the other hand, M3D technology uses a single base wafer to include components from different layers, avoiding problems related to alignment, thinning, and TSV bonding [10]. This integration of similar chip layers creates a unified three-dimensional chip structure. Ultimately, M3D improves how chips are placed in a limited space and enables quick connections, making chip integration more efficient and allowing electronic systems to operate faster.

## 3 PROPOSED METHODOLOGY

### 3.1 Motivation

Accelerators play a crucial role in executing convolutional and fully-connected layers, which involve numerous matrix multiplications. Typically, arrays of multipliers in the accelerators perform these multiplications. However, a one-size-fits-all fixed architecture does not guarantee optimal performance. The best accelerator architecture varies depending on factors such as input size, model size, and the number of layers. To exemplify this point, let us consider a scenario where we have a 64x64 systolic array and three different CNN model inference tasks to be executed on it. Initially, all three models are executed on the unified array. Subsequently, the unified array is decomposed into four arrays of size 32x32, positioned in 2x2, 4x1, and 2x1H arrangements, as depicted in Fig. 2 (a), (b), and (c) respectively. In this example, the networks are ResNet50 [11], OCR [12], and Speaker ID [13].

Fig. 3 demonstrates the number of execution cycles required for each of these networks in different arrangements, normalized to the minimum value obtained for each network. Interestingly, the best arrangement differs for each network. For the OCR network, the 4x1 (linear) arrangement yields the best execution cycle, with a performance improvement of over two-fold over the worst one. Notably, this arrangement increases the execution cycle of the ResNet50 network by almost 30% compared to its best arrangement (2x2). For the Speaker ID network, we observe that the 2x1H arrangement provides the best outcome. This specific arrangement also yields the worst result for the ResNet50 network and an average result for the OCR network.

In summary, the motivation behind this research stems from the observation that fixed accelerator architectures do not always deliver optimal performance across various CNN models. Moreover, different network configurations exhibit diverse performance outcomes due to the mapping of neural network layers onto computational resources and the subsequent variation in utilization percentages. Therefore, an imperative need emerges to develop reconfigurable accelerators to maximize performance across different networks.

### 3.2 The Multi-Objective Problem and Design Space Exploration Framework

As we will show in section 3.3, the proposed architecture allows for reconfigurability among different scale-up and scale-out arrangements to obtain the best trade-off across power, energy, and performance when accelerating CNN inference. In order to effectively explore the state space of design parameters, a design space exploration framework is introduced in this section. To this end, we first present a multi-objective problem aimed at optimizing the execution of CNN inference on systolic arrays. The primary objectives are to minimize the number of execution cycles and increase utilization, while secondary goals include minimizing power consumption, energy consumption, and memory access counts. Extensive experiments are conducted using this framework to obtain desired values of the fixed parameters of the reconfigurable architecture.

The proposed architecture is implemented using M3D technology in four layers. However, users will have the flexibility to determine modifications and arrangements relevant to neural networks, such as the model size and input data. The CNN inference accelerator architecture is influenced by various determining parameters, including the type and number of processing elements, operating frequency, memory architecture, base array size, data flow, arrangements (scale-out vs. scale-up), etc. These parameters have a direct impact on the utilization and performance of the accelerator.

The aim is to thoroughly search the design space of the multi-objective problem to tune the control design parameters. A framework is provided to conduct experiments by assigning values to these control parameters. Without loss of generality, we assume the control design parameters are the base array size, dataflow, and systolic array scale-out arrangements. Among these, the first two are tuned in the design space exploration step, while the last one (arrangements) will be used as the reconfigurable parameter. The complete steps of the framework are depicted in Fig. 4, with corresponding descriptions provided for each number presented in the figures.

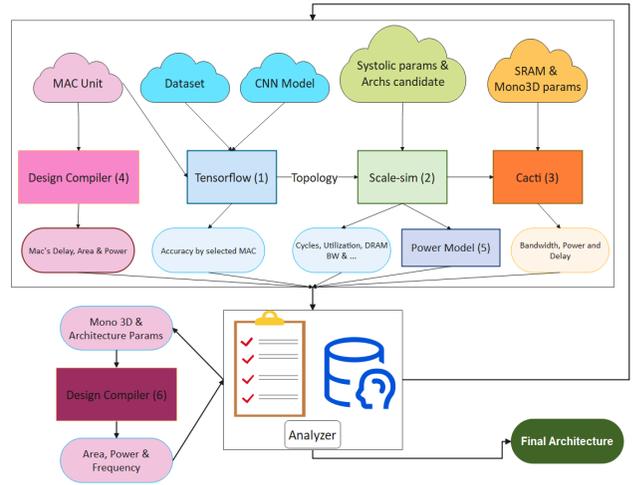

Fig. 4. Design space exploration framework.

Step (1) in the process encompasses the application of TensorFlow [14] to implement neural network models. The structure of the neural network models is specified and subsequently exported to the SavedModel Format. A trained frozen model, along with its associated graphs and computations, is acquired to facilitate the simulation of model inference within the SCALE-Sim [15], a cycle-accurate simulator. Step (2) utilizes the SCALE-Sim simulator to simulate systolic arrays, receiving the network topology and architecture configuration as input. It delivers various execution cycle factors for each layer, processing element utilization, and memory accesses, among others, upon running the neural network on the processing resources. Step (3) involves the utilization of the CACTI [16] tool to implement and simulate the behavior of the cache. The results from running the neural network on the processing resources, which encompass memory accesses, allocated memory space, memory type, etc., serve as input for the CACTI tool. The simulator then extracts the power consumption, cycle delay, and other related factors.

Step (4) of the procedure involves the synthesis of the Multiply-Accumulate (MAC) unit using the Design Compiler [17] tool, utilizing the $15nm$ NanGate technology library [18] for the synthesis and implementation process. The resulting synthesized MAC unit is subsequently employed as input in step (5), where a power consumption calculation model [19] is utilized to determine the power consumption of the systolic arrays. Following the analysis of the obtained results and performing design space exploration, the final accelerator architecture is determined. This proposed architecture is then implemented at the register-transfer level using SystemVerilog in step (6) to estimate delay, power, and area.

**Algorithm 1:** Design Space Exploration Procedure
**Input:** Neural Network Model Set ($M$), Systolic Base Size Set ($B$), Arrangement Set ($A$)
$D$ = Data flow set = {OS, WS, IS}
$B$ = Systolic Base Size set = {32, 64, …}
$A$ = Arrangement = {1x1, 2x1, …}
$M$ = NN model set = {FaceRecognition, …}
$C = D \times B$
$S = C \times A$
**Output:** Architecture parameter values ($c^*$)
1  **begin**
2    *# Design space pruning based on thresholding.*
3    Remove $\alpha$% of $S$ with the worst design parameter values.
4    **foreach** ($m \in M$) **do**
5      **foreach** ($s \in S$) **do**
6        $O_m$ (s, : ) = [delay, power, utilization, memory access]
7      **endfor**
8      $P_m$ = Find pareto front set of $O_m$
9      **foreach** ($s \in S$) **do**
10       $d_m(s)$ = Find distance of $O_m(s)$ to $P_m$ = $\min_{p \in P_m} \| O_m(s) - p \|_2$
11     **endfor**
12   **endfor**
13   **foreach** ($c \in C$) **do**
14     **foreach** ($m \in M$) **do**
15       **foreach** ($a \in A$) **do**
16         $r(c) = [r(c),\ d_m(s = (c,a))]$
17       **endfor**
18     **endfor**
19     $r(c)$ = average($r(c)$)
20   **endfor**
21   **return** $c^* = argmin(r(c))$
22 **end**

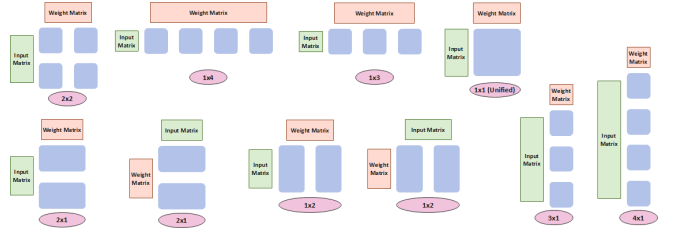

Fig. 5. Systolic array arrangements.

The execution of each model is classified into three main categories, corresponding to the general data flow states. This implies that the neural network model is executed for Output-Stationary (OS), Weight-Stationary (WS), and Input-Stationary (IS) states. Subsequently, for the selected data flow scaling, a decision is made regarding whether the systolic array should be scaled out or scaled up, which is further divided into multiple arrangements. The main concern here is whether the systolic array should be utilized as a whole or divided into smaller systolic arrays with an equal number of processing elements. Since the improvement resulting from the division is not certain for distinct models, experimentation should be conducted for both cases.

Hence, the configurations of systolic arrays can be categorized into unified alternatives and partitioned into two, three, or four sections. However, dividing the array beyond a certain threshold and surpassing a specific limit for the number of arrays results in a considerable strain on the communication network and diminishes processing capability. Upon dividing the systolic array, two modes emerge that necessitate a determination regarding whether the input matrix should be entered gradually from the top (in compliance with the dataflow of systolic array data) or from the left. These conditions similarly hold true for the weight matrix. When applying these two mapping modes to the aforementioned unified and partitioned systolic array alternatives, ten distinctive arrangements can be derived, as depicted in Fig. 5.

Thus far, decisions have been made regarding different data streams and the mapping of input matrices. However, the number of processing elements to be included in each systolic array base has not been discussed. To determine the base value, this variable is altered within the range of 32 to 512 through various experiments to determine the optimal value. Beyond the size of 512, the utilization becomes nearly zero, and reducing the size below 32 significantly increases the number of execution cycles. To observe the effects of different scenarios on various workloads, a range of workloads must be applied. For instance, an architecture may perform exceptionally well for heavy workloads but yield unfavorable outcomes for lighter workloads. As mentioned earlier, a fixed architecture may not be optimal for all scenarios. In order to assess various workloads, this study utilizes a set of four neural networks as benchmark models, categorized as light, medium, heavy, and very heavy workloads, each aligned with specific network architectures such as Face Recognition [20], DeepSpeech [21], ResNet50, and AlexNet [22], respectively.

The proposed design space exploration algorithm (Algorithm 1), aims to explore the design space of neural network architectures for efficient hardware implementation. The algorithm takes as input a set of neural network models ($M$) and a design space ($S$), which is defined as the Cartesian product of the data flow set ($D$), the systolic base size set ($B$), and the arrangement set ($A$). The design space is pruned by removing $\alpha$% of the solutions with the worst design parameter values, ensuring a more focused exploration.

For each neural network model ($m$) in the model set, the algorithm evaluates each solution ($s$) in the design space based on four key design parameters: delay, power consumption, resource utilization, and memory access. Subsequently, a Pareto front set ($Pm$) is determined for each model to identify non-dominated solutions in the multi-objective optimization space. The algorithm then computes the distance of each solution in the design space to its respective Pareto front, using the Euclidean norm. Finally, an average distance metric ($r(c)$) is computed for each solution across all neural network models and systolic arrangements. The optimal solution ($c^*$) is identified as one that minimizes this average distance metric, effectively providing the architecture configuration (base size and data flow) that balances the trade-off across the considered four design parameters.

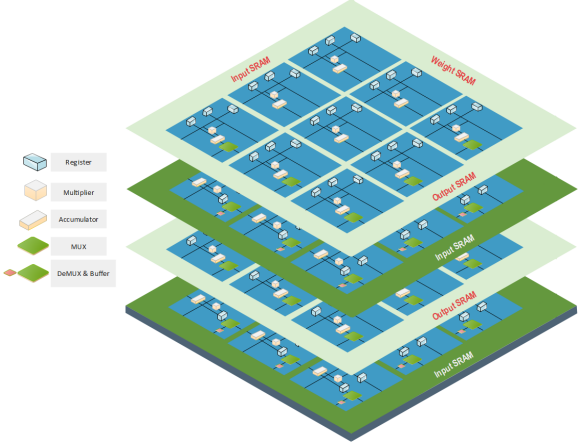

Fig. 6. The ARMAN architecture.

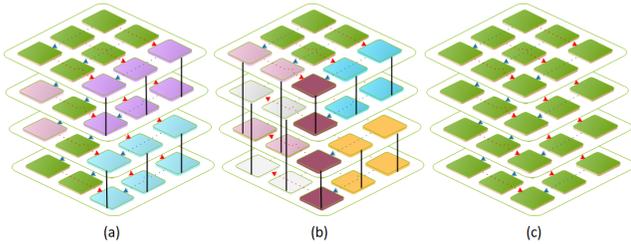

Fig. 7. Connections in three different accelerator arrangements a) 2x1 scale-out, b) 1x1 scale-up, and c) 2x2 scale-out.

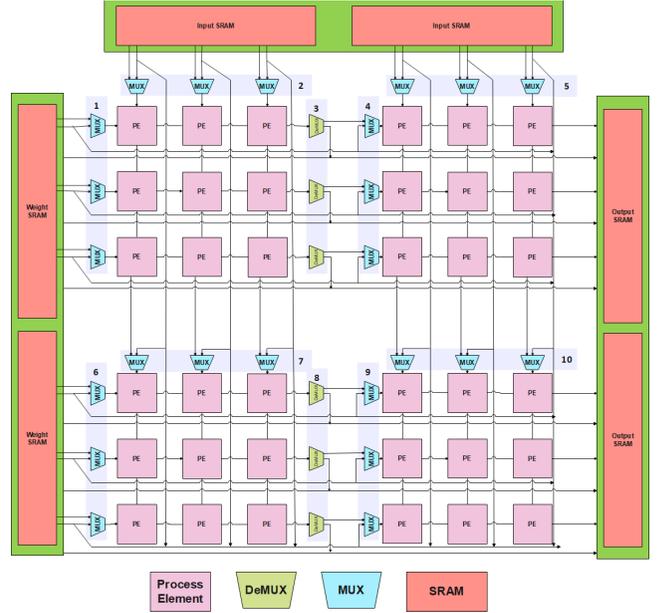

Fig. 8. Reconfigurable interconnect network of the proposed architecture.

The findings were derived under the presumption that certain points were distributed randomly according to the threshold level ($\alpha$). Various thresholds were assessed and ultimately set at 30%. Based on the observed outcomes, the optimal dataflow strategy is OS, with WS following closely behind and exhibiting a moderate divergence. Additionally, the most effective base size value for the accelerator architecture was identified as 64, with 128 as the second-best option and a noticeable gap between them. Section 4 provides a detailed examination of the results associated with the aforementioned cases.

### 3.3 Accelerator Architecture

In this section, we present ARMAN, the proposed reconfigurable accelerator architecture for CNN inference. Our proposal involves a configurable systolic array-based accelerator that can dynamically switch between different scale-out arrangements based on the input workload. The accelerator includes separate input/output buffers and control logic for each mode, along with a mechanism to switch between modes based on input dataflow. Fig. 6. provides a high-level view of the architecture, highlighting the utilization of M3D for interconnecting accelerator layers. Additionally, each row of the accelerator contains a multiplier array, with input, weight, and output cache memories placed next to it in each layer.

The base architecture used for processing elements in each layer is a systolic array, enabling parallel and simultaneous convolution operations, resulting in increased execution speed and reduced execution time. With data stored locally and close to the computational units, memory traffic and access to external memory are minimized. Moreover, the architecture leverages the smallest computational units for convolution computations and facilitates data movement within the array. As a result, power consumption decreases, and scalability and expandability are enhanced, enabling the processing of larger neural networks.

Each systolic array consists of 64x64 processing elements, with each processing element able to send its data to the next processing element in the same layer (located to the right or down) or to the processing element in the next layer located below it. By default, each systolic array operates separately in a scale-out manner. However, a scale-up mode can be configured, where all systolic arrays have data transfer in a way that they function as an integrated (unified) systolic array. A unified 128x128 systolic array can be constructed through the combination of four layers. This is achieved by first integrating two small square-shaped systolic arrays, resulting in rectangular-shaped systolic arrays. Subsequently, the rectangular systolic arrays obtained from the previous step are merged to create a unified systolic array. It is important to highlight that these interconnections are established using Monolithic Inter-tier vias (MIVs) within the three-dimensional architectural framework. The implementation of M3D technology allowed us to successfully integrate a cache memory within each systolic array. By adopting this approach, we were able to minimize and equalize the critical path connecting each systolic array to the cache memory. Alternatively, an intermediate mode allows for merging two systolic arrays to form a 2x1 or 1x2 arrangements, to create two systolic arrays of 64x128 or 128x64. Fig. 7 showcases these different arrangements.

The interconnect network of the accelerator, represented by a 3x3 array in Fig. 8, allows for reconfigurability. The memory banks have two output ports, enabling reconfiguration through this interconnect network. The network utilizes a demultiplexer in the source processing element and a multiplexer in the destination processing element for reconfiguration purposes. The Mux and De-Mux selectors are shown in Table 1 for different arrangements. The columns are the index of the Mux/DeMux group as shown in Fig. 8.

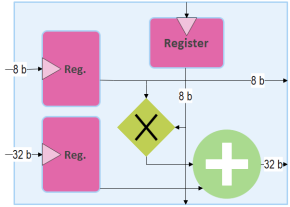

Fig. 9. MAC unit structure.

TABLE 1 MUX AND DEMUX SELECTORS

| Arrangement | 1 | 2 | 3 | 4 | 5 | 6 | 7 | 8 | 9 | 10 |
|---|---|---|---|---|---|---|---|---|---|---|
| 2x2 | 1 | 1 | 1 | 1 | 1 | 1 | 1 | 1 | 1 | 1 |
| 1x4 | 1 | 0 | 1 | 1 | 0 | 1 | 1 | 1 | 1 | 1 |
| 4x1 | 0 | 1 | 1 | 1 | 1 | 0 | 1 | 1 | 1 | 1 |
| 1x3 | 1 | 0 | 1 | 1 | 0 | - | - | - | 1 | 1 |
| 3x1 | - | - | - | 1 | 1 | 0 | 1 | 1 | 1 | 1 |
| 1x2 | 1 | 0 | 1 | 1 | 0 | 1 | 0 | 1 | 1 | 0 |
| 2x1 | 0 | 1 | 0 | 0 | 1 | 0 | 1 | 0 | 0 | 1 |
| 1x1 | 0 | 0 | 0 | 0 | 0 | 0 | 0 | 0 | 0 | 0 |

We employed an 8-bit width MAC unit as the basic unit for performing inference, similar to Google TPU architecture (Fig. 9) [23].

After exploring the design space using Algorithm 1 and examining the results obtained, an OS data flow is adopted in the accelerator architecture (as we will show in detail in section 4). This data flow allows for increased parallelism and requires less memory access compared to IS or WS arrays. To store weights, activations, and intermediate results of CNN layers, an on-chip cache memory is utilized in this architecture. This reduces off-chip memory access and facilitates faster data access. Off-chip DRAM is employed to store larger datasets, functioning as a buffer between the on-chip cache memory and the host central processing unit. This enables the processing of larger datasets and facilitates data transfer to and from the central processing unit.

## 4 EXPERIMENTAL RESULTS

We perform design space exploration and assess the performance, power and energy consumptions of the architectures on four distinct AI workloads, i.e., Face Recognition, DeepSpeech, ResNet50, and AlexNet. The experiments were conducted on systolic arrays of five different base sizes: 32, 64, 128, 256, and 512, for three different data flows: OS, WS, and IS, and for ten different arrangements: 1x1, 2x2, 4x1, 1x4, 1x3, 3x1, 1x2H, 1x2V, 2x1H and 2x1V. We first provide the analysis of the data gathered by running these workloads under different configurations. Then, we provide the result of running the design space exploration algorithm to extract reconfiguration candidates. Finally, we present the evaluation of the proposed reconfigurable architecture.

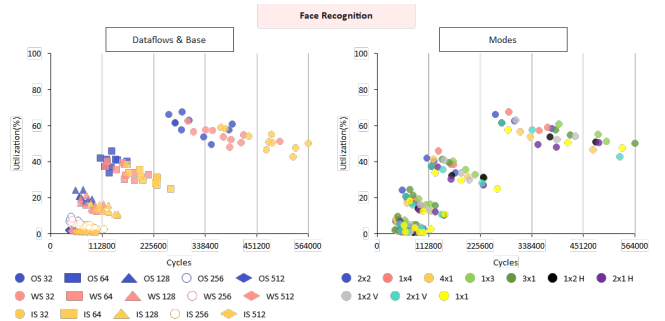

Fig. 10. The number of cycles and utilization values for all systolic array configurations under the Face Recognition workload.

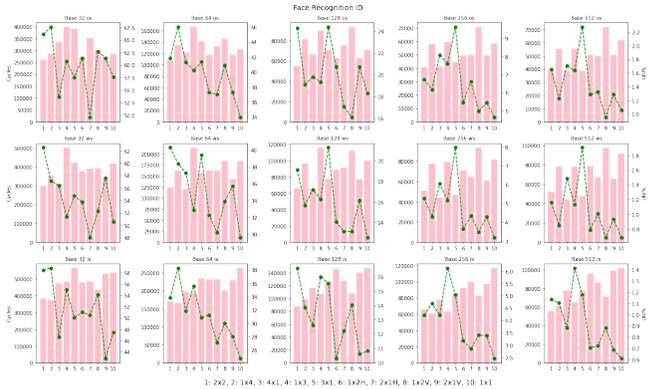

Fig. 11. The Face Recognition workload was categorized by base and dataflow, with corresponding values for the number of cycles and utilization.

### 4.1 Workload Analysis

#### 4.1.1 Face Recognition

We select this network as a representative example for light-weight workloads (in terms of the number of computations). Figs. 10 and 11 show the execution cycles and utilization of this network over the range of different base sizes, data flows, and arrangements. The OS dataflow was found to consistently provide the best results in terms of execution cycles and utilization for almost all base sizes. Also, the decrease in the execution cycle and the utilization of processing elements is evident from Fig. 10 when increasing the base size of the systolic arrays. The distances between the execution cycles become closer as the base size increases, indicating a diminishing return in execution cycle (from over 100% when moving from 32 to 64, to below 10% when moving from 256 to 512 base size). The utilization rates were found to be 60%, 40%, 20%, 5%, and 1% on average for base sizes 32, 64, 128, 256, and 512, respectively for OS data flow. Hence, further increases in base size over 64 would not be beneficial for this light-weight network, as the utilization decreases below 25%.

Furthermore, as depicted in Fig. 11, the 2x2 arrangement consistently yields the best performance for this network. It can be inferred that the second-best arrangement is the 4x1 one. Changing the arrangement to other cases may decrease the execution cycles significantly, e.g., up to 40% for 64 base size and OS data flow.

Fig. 12 presents the power and energy diagrams for this network under different configurations. It is observed that as the array size increases, the power consumption also increases due to the increase in the number of processing elements. However, the energy consumed remains relatively constant across different base sizes. This is attributed to the fact that although the power consumption of systolic arrays increases with array size, the number of execution cycles decreases, resulting in a nearly constant energy consumption.

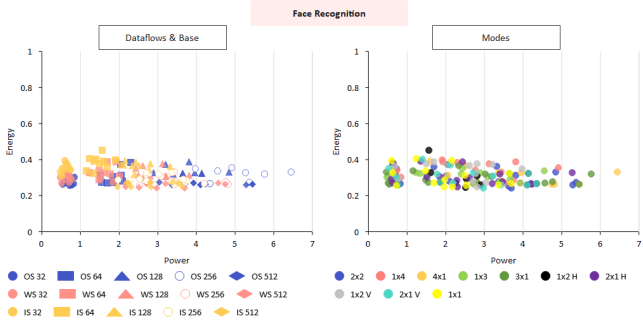

Fig. 12. Power and energy for all systolic array configurations under the Face Recognition workload.

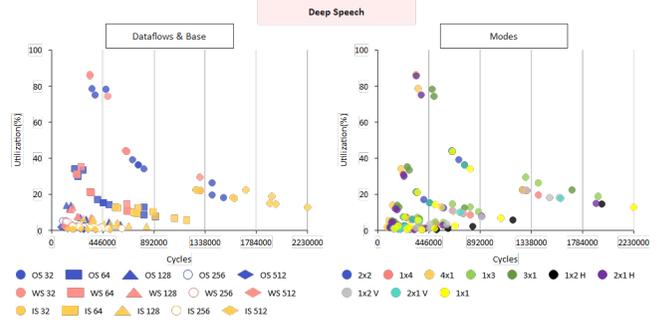

Fig. 14. The number of cycles and utilization values for all systolic array configurations under the DeepSpeech workload.

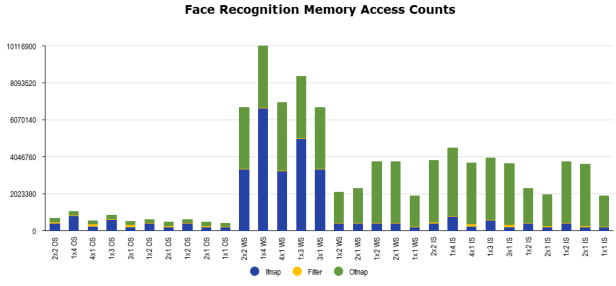

Fig. 13. Memory access count under the Face Recognition workload.

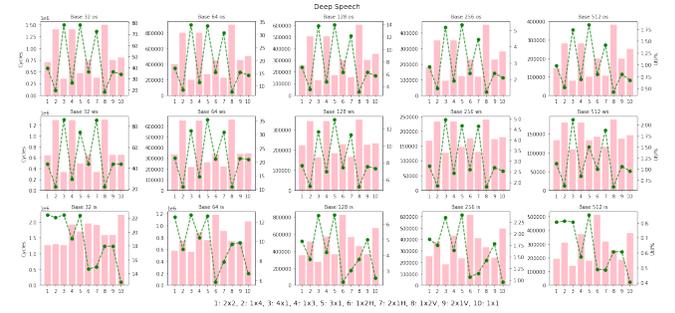

Fig. 15. The DeepSpeech workload was categorized by base and dataflow, with corresponding values for the number of cycles and utilization.

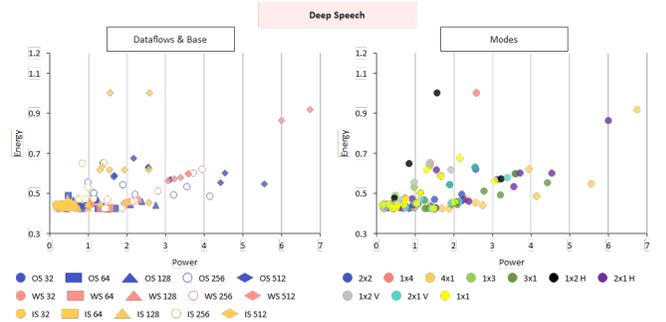

Fig. 16. Power and energy for all systolic array configurations under the DeepSpeech workload.

The analysis of the results further reveals that, for cases where the accelerator needs to operate at a lower power state, the optimal arrangements are 3x1 and 1x3 due to their lower number of MAC units compared to other arrangements. Interestingly, the energy consumption is lower for OS data flow for base sizes 32 and 64, but the WS and IS data flows show better energy consumption than OS when the base size increases above 64.

Fig. 13 illustrates the memory access counts for base size of 64. The results show that the OS dataflow exhibits a notable improvement in memory access compared to alternative data flows. Moreover, within the various arrangements tested, the unified 1x1 arrangement demonstrates the least number of memory accesses, followed by 2x1, 3x1, and 4x1 arrangements.

**Summarized observations:** The results reveal that the OS dataflow has exhibited superior performance across all parameters for this network. In terms of cycle count and utilization, the 2x2 and 4x1 arrangements have demonstrated the most favorable results. Conversely, the 3x1 and 1x3 arrangements have proven to be the most efficient in terms of power consumption and energy usage. Additionally, the 1x1, 2x1, 3x1, and 4x1 configurations have excelled in terms of memory access. Furthermore, the presence of 32 and 64 bases has been determined to be the optimal number of MAC units for benchmarking this light-weight network.

### 4.1.2 DeepSpeech

We select DeepSpeech as a representative example of medium workloads. Figs. 14 and 15 show the execution cycles and utilization of this network over the range of different base sizes, data flows, and arrangements. The results indicate that the utilization rates are 80%, 35%, 18%, 4%, and 1% on average, respectively for base sizes of 32 to 512. In addition, the utilization does not exceed 30% for IS data flow, rendering it inappropriate for this workload. However, OS and WS data flows show comparable performance over the range of different base sizes. Interestingly, the 32 base size demonstrates above 80% utilization for 2x1H arrangement, but as shown in Fig. 15, the utilization may decrease down to 40% by using 1x1 unified arrangement (for OS data flow). The execution cycle also decreases by 3x by only changing the arrangement for this base size and data flow. These observations emphasize the significance of choosing the right arrangement and dataflow for optimal performance. Furthermore, the number of cycles did not decrease significantly above size 128. In summary, the optimal configurations in terms of performance are the 4x1 and 2x1H arrangements and OS data flow.

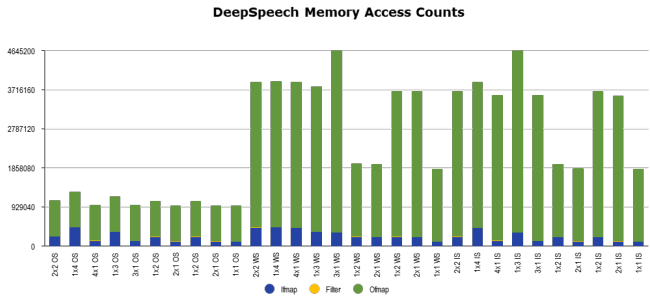

Fig. 17. Memory access count under the DeepSpeech workload.

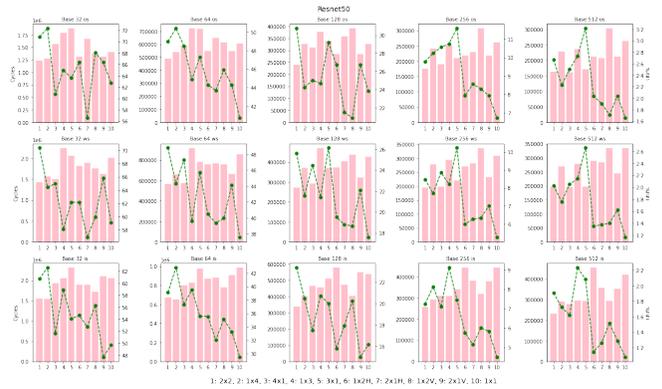

Fig. 19. The ResNet50 workload was categorized by base and dataflow, with corresponding values for the number of cycles and utilization.

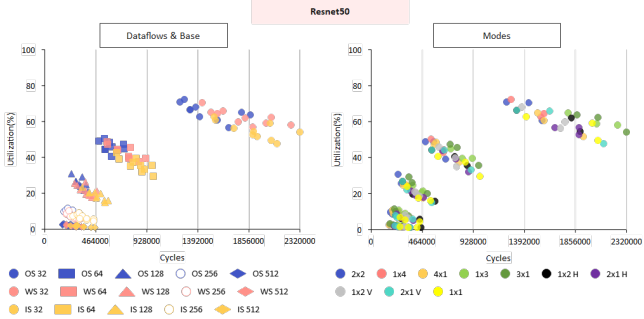

Fig. 18. The number of cycles and utilization values for all systolic array configurations under the ResNet50 workload.

Fig. 16 displays the power and energy consumptions of DeepSpeech network, highlighting that the IS data flow shows improved power consumption over OS and WS ones. In the OS and WS data flows, the 1x3, 1x4, and 1x2V arrangements exhibit lower power consumption over different base sizes. However, the best arrangement in terms of energy consumption differs from one base size to another, e.g., 1x1 is the best one for base size of 32, but 2x1H is the best for base size of 64, and so on.

Fig. 17 illustrates that the OS dataflow exhibits lower number of memory accesses relative to other data flows. Furthermore, the 1x1 arrangement is also identified as the most favorable one in terms of memory access count, in the base size of 64.

**Summarized observations:** In summary, the OS dataflow has demonstrated optimal performance across all parameters. Among the various arrangements tested, the 2x1H and 4x1 configurations have exhibited superior cycle count and utilization. Alternatively, the 1x3, 1x2V, and 1x4 arrangements have excelled in terms of power consumption. Moreover, the 1x1 arrangement has proven to be the most efficient in memory access. The employment of 32 and 64 base sizes has yielded the best results for this network.

### 4.1.3 ResNet50

The ResNet50 network is considered to represent semi-heavy workloads. As illustrated in Fig. 18, increasing the base size from 128 onwards only led to marginal performance improvement before reaching a point of saturation. Additionally, it is observed that increasing the base size from 32 to 64 has had a significant positive impact on the number of cycles. Moreover, at the base size of 64, the utilization of this network is approximately 50%, and again, the OS data flow surpasses other data flows.

Fig. 19 shows that the optimal arrangement is 2x2 over the range of different base sizes. This particular network has reached a utilization rate of almost 30% for base size of 128 since it is more compute-intensive than the previous networks. Changing the arrangement to other cases may decrease the execution cycles significantly, e.g., up to 30% for 64 base size and OS data flow.

As indicated in Fig. 20, the energy and power levels demonstrate almost 20% and 10x decrease when decreasing the base size. Hence, base size of 32 exhibits comparatively lower power consumption than other ones. Moreover, the results suggest that the OS data flow shows improved energy consumption similar to Face Recognition network. In the OS data flow, the optimal configurations, in terms of power consumption, are 3x1 and 1x3 for base sizes of 32-128, and 1x1 for base sizes of 256-512.

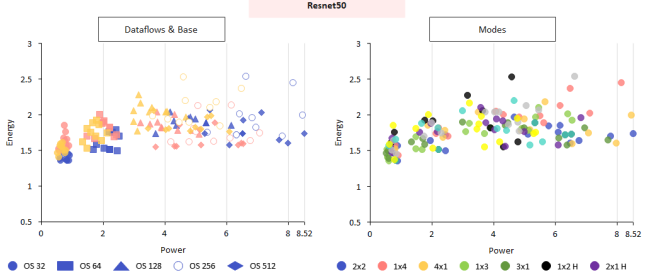

Fig. 20. Power and energy for all systolic array configurations under the ResNet50 workload.

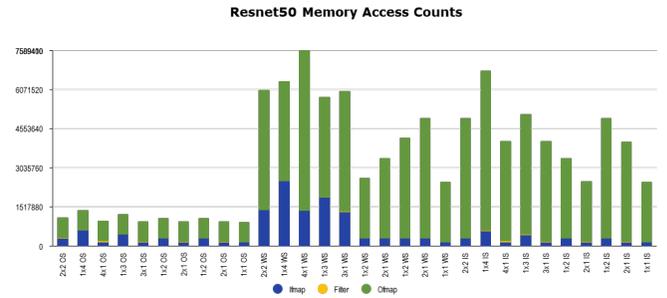

Fig. 21. Memory access count under the ResNet50 workload.

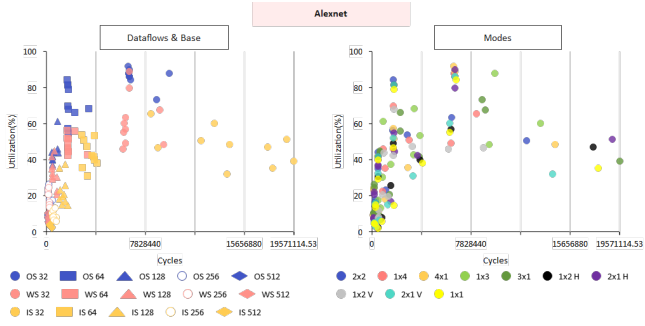

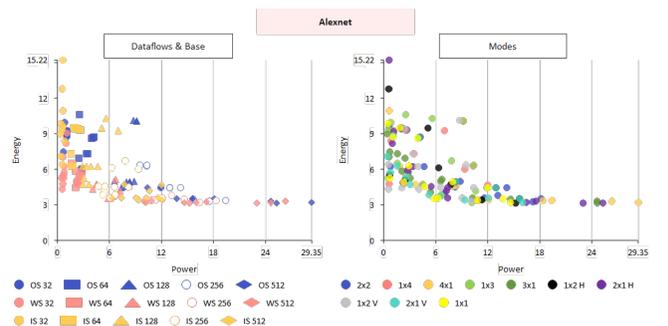

Fig. 22. The number of cycles and utilization values for all systolic array configurations under the AlexNet workload.

Fig. 24. Power and energy for all systolic array configurations under the AlexNet workload.

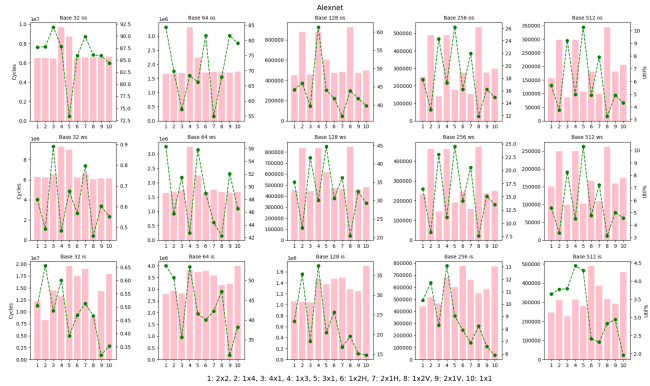

Fig. 23. The AlexNet workload was categorized by base and dataflow, with corresponding values for the number of cycles and utilization.

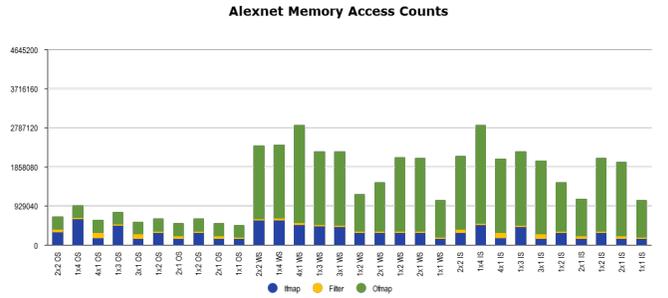

Fig. 25. Memory access count under the AlexNet workload.

Fig. 21 shows that the OS data flow and 1x1 arrangement again exhibit the optimal number of memory accesses.

**Summarized observations:** The OS data flow has exhibited superior performance across all configurations for this particular workload. The 2x2 arrangement has proven to provide the best performance. In terms of energy consumption, the 3x1, 1x3, and 1x1 arrangements have demonstrated the most favorable results. Additionally, the integrated 1x1 arrangement exhibits advantageous memory access count.

*4.1.4 AlexNet*

The AlexNet network is characterized by its intensive computational demands, as evidenced by Fig. 22 and Fig. 23. Compared to previous networks, this network is exceptionally large, demanding much more computation power. Fig. 22. reveals the performance superiority of OS data flow to the other ones in this workload type, similar to previous results, i.e., it provides much lower execution cycles by reaching higher utilization ratios. Additionally, as the base size increases, the utilization drops substantially as expected. It is important to note that the decrease in cycle count indicates a saturating trend in speedup as we increase the base size. However, due to the large size of the network, it still benefits much from very large base sizes (256 and 512). More specifically, the speedup starts from 60x when moving from 32 to 64 base size, and reaches 1.5x as we move from 256 to 512. The network's utilization also drops below 5 percent at the base size of 512. The 1x2V, 2x2 and 4x1 arrangements provide the best performance in base sizes of 64, 128 and 256.

Fig. 23. shows that for the OS data flow, the execution cycle count is almost similar for all arrangements except for 1x3 and 3x1 for base sizes of 32 and 64. However, for base sizes of 256 and 512, 4x1 and 2x1H provide substantially better cycle count.

Fig. 24. demonstrates that WS data flow offers the best power-energy profile. Notably, the optimal arrangement concerning power consumption in this benchmark network is 1x2V.

In Fig. 25. the OS data flow shows improved memory access count compared to other alternatives which is consistent with previous observations. Among various arrangements in this data flow, the 1x1 arrangement is identified as the best one, followed by 2x1H.

**Summarized observations:** The OS data flow has exhibited superior performance across all configurations for this particular workload. The 2x2 arrangement has proven to provide the best performance. In terms of energy consumption, the 3x1, 1x3, and 1x1 arrangements have demonstrated the most favorable results. Additionally, the integrated 1x1 arrangement exhibits advantageous memory access count

### 4.2 Evaluation

Using all the data gathered in the workload analysis step, we run Algorithm 1, a Pareto-based design space exploration procedure, to find the optimal configuration that yields the best trade-off across our four design goals. Table 2 shows the optimal configurations obtained under different values of the pruning threshold ($\alpha$).

Using the threshold values, the overall best configuration emerges to be OS data flow and base size of 64. Although using other values of $\alpha$ may lead to different optimal configurations, we proceed with the rest of evaluations using this choice. The outcomes of the conducted synthesis can be observed in Table 3. Note that the clock frequency is degraded only by 3.3% due to the introduction of Mux/DeMuxs in the critical path, which is negligible.

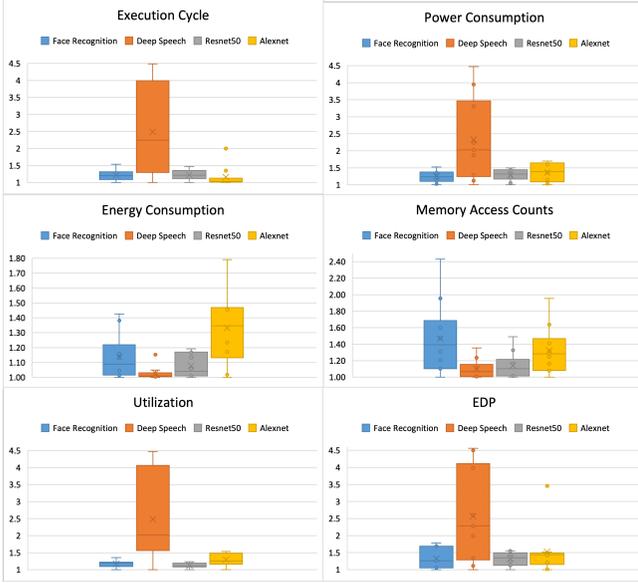

Fig. 26. Evaluation of the proposed architecture under four benchmarks. The results are gathered across different arrangements.

TABLE 2 OPTIMAL CONFIGURATIONS FOR DIFFERENT THRESHOLDS

| No | Threshold ($\alpha$) | | | | | |
|---|---|---|---|---|---|---|
| | 20% | 25% | 30% | 35% | 39% | 40% |
| 1 | OS-64 | OS-64 | OS-64 | OS-64 | OS-64 | OS-64 |
| 2 | WS-256 | WS-256 | OS-128 | WS-128 | WS-128 | |
| 3 | OS-128 | WS-128 | WS-128 | IS-128 | | |
| 4 | WS-128 | OS-128 | IS-128 | | | |
| 5 | IS-256 | IS-128 | | | | |
| 6 | WS-64 | WS-64 | | | | |
| 7 | IS-128 | | | | | |

TABLE 3 SYNTHESIS RESULTS

| Accelerator parameters | Values |
|---|---|
| Clock rate (reconfigurable) | 800 MHz |
| Clock rate (non-reconfigurable) | 826 MHz |
| Number of PEs | 128x128 |
| Number of layers | 4 |
| Multipliers bit width | 8 |
| Cache size | 8 MB |
| Technology | 15nm FinFET |
| Area (reconfigurable) | 324.1 $mm^2$ |
| Area (non-reconfigurable) | 322.05 $mm^2$ |

Fig. 26. shows the box plots of the six design parameters, for the OS data flow and base size of 64. The values are illustrated across different arrangements. The results confirm that reconfiguring the arrangement to the best one can achieve up to almost 1.55x, 4.5x, 1.45x, and 2x improvements in execution cycle, for Face Recognition, DeepSpeech, ResNet50, and AlexNet models, respectively. The improvement in power consumption is almost similar to execution cycle, with the exception of AlexNet, which is 1.7x. Moreover, the reconfiguration of arrangements can lead up to 1.42x, 1.15x, 1.2x, and 1.8x energy improvement for the mentioned models respectively.

Finally, the energy-delay-product can be improved by up to 1.8x, 4.6x, 1.7x, and 3.5x respectively. These results verify that our reconfigurable architecture is able to largely improve each design parameter when necessary, by rearranging the systolic array to the corresponding best arrangement for that design parameter. Note that these improvements are reported only for the computation-intensive parts of the CNNs, since SCALE-Sim does not support other non-compute-intensive parts such as activation and batch normalization.

It is important to note that the best arrangement differs for each benchmark and each design parameter. For example, the best arrangement in terms of execution cycle is 4x1 for DeepSpeech, and 2x2 for the other workloads. However, in terms of power consumption, the best arrangement is 1x3 for FaceRecognition and ResNet50, 1x2H for DeepSpeech, and 3x1 for AlexNet. For energy consumption, 2x2, 2x1V, and 1x1 are optimal in different cases. Table 4 gives these best arrangements for each benchmark and each design parameter. The table also shows the corresponding improvements with respect to the fixed 2x2 arrangement, as a widely used scale-out arrangement. These results verify that different workloads require different arrangements to reach optimal performance, power, energy, or EDP, and the proposed reconfigurable architecture can effectively exploit this potential to improve the objectives. In summary, the execution cycle, power, energy, and EDP improve by up to 2x, 2.24x, 1.48x, and 2x respectively for our selected networks.

TABLE 4 OPTIMAL ARRANGEMENTS AND CORRESPONDING IMPROVEMENTS W.R.T. FIXED 2X2 ARRANGEMENT

| Benchmark | Cycle | Imp. | Power | Imp. | Energy | Imp. | EDP | Imp. |
|---|---|---|---|---|---|---|---|---|
| Face Recognition | 2x2 | 1x | 1x3 | 1.39x | 1x1 | 1.05x | 2x2 | 1x |
| DeepSpeech | 4x1 | 2x | 1x2H | 2.24x | 2x1V | 1.01x | 4x1 | 2x |
| ResNet50 | 2x2 | 1x | 1x3 | 1.46x | 2x2 | 1x | 2x2 | 1x |
| AlexNet | 2x2 | 1x | 3x1 | 1.7x | 2x1V | 1.48x | 4x1 | 1.42x |

## 5 RELATED WORK

This section aims to present an overview of the research endeavors in accelerator architectures, emphasizing noteworthy advancements and approaches in this dynamic domain.

Schokla et al. [20] proposed a temperature-conscious optimization framework for 3D-integrated deep neural network accelerators, addressing challenges while maintaining efficiency. This involves critical pathway identification, thermal behavior modeling, and energy-efficient layer mappings within temperature constraints.

Yu et al. [24] developed the SPRING architecture using reduced precision and sparse matrix techniques to create an efficient 3D CNN accelerator. Their approach outperforms existing architectures in terms of efficiency and energy usage by strategically using approximate computations.

Joseph et al. [25] conducted a thorough analysis of the benefits and challenges of 3D technology in deep neural network accelerators, offering solutions like microfluidic cooling and specialized interlayer connections. Their innovative work introduces a 3D design methodology that overcomes the limitations of traditional 2D circuits.

Arnand et al. [26] proposed a strategy to improve the efficiency and energy economy of the GEMM accelerator by integrating machine learning. Their approach combines the accelerator with a framework that learns mappings between settings and performance outcomes, enhancing energy gains and efficiency.

Lim et al. [27] developed the FlexSA systolic array architecture for reduced-precision deep neural network training, enhancing speed and efficiency through adaptability, weight pruning, and tensor handling. The distinct features of FlexSA constitute the core innovation of their work.

Tasoulas et al. [28] developed an energy-efficient hardware accelerator for neural network inference by using weight-based approximations, effectively balancing energy savings and accuracy. If this approach can be applied to other neural networks, it can potentially address limitations that arise with larger network dimensions.

Yuzguler et al. [29] investigated the impact of network topology on training efficiency in distributed deep learning systems. They developed an automated tool to identify optimal topologies based on experiments conducted on various models.

Table 5 presents an overview and comparative analysis of the related work and the proposed architecture.

TABLE 5 RELATED WORK OVERVIEW

| NO. | Work | 3D Technology | Systolic Array | Reconfigurable | Power |
|---|---|---|---|---|---|
| 1 | Schokla et al [19] | ✓ | ✓ | ✗ | ✓ |
| 2 | Yu et al [24] | ✓ | ✗ | ✗ | ✓ |
| 3 | Joseph et al . [25] | ✓ | ✓ | ✗ | ✗ |
| 4 | Arnand et al [26] | ✗ | ✓ | ✓ | ✓ |
| 5 | Lim et al [27] | ✗ | ✓ | ✗ | ✗ |
| 6 | Tasoulas et al [28] | ✗ | ✓ | ✓ | ✓ |
| 7 | Yuzguler et al [29] | ✗ | ✓ | ✗ | ✓ |
| 8 | ARMAN (this work) | ✓ | ✓ | ✓ | ✓ |

## 6 CONCLUSION

Using hardware accelerators can significantly enhance the performance, power, and energy consumption of CNN inference. However, a fixed-architecture accelerator for inference of CNN models may not consistently yield the best results for all neural networks. The optimal architecture can vary depending on design objective and also the model attributes such as input, size, and number of layers in the model. To address this challenge, a reconfigurable M3D systolic-array-based architecture is proposed. Throughout extensive experiments on four distinct NN models, we showed that the proposed architecture can effectively enhance performance, power, energy, and energy-delay-product of the accelerator.

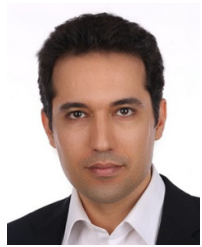

**Mahmoud Momtazpour** received his B.Sc., M.Sc., and Ph.D. degrees in electrical engineering from Sharif University of Technology, Iran, in 2005, 2007 and 2012 respectively. He is currently an Assistant Professor with the Department of Computer Engineering, Amirkabir University of Technology, Iran. His current research interests include system-level performance and energy optimization in parallel and distributed systems, including cloud-based systems, Internet of things, and GPU/multicore-based systems.

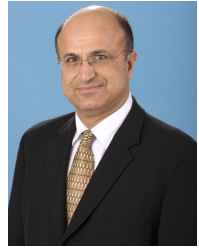

**Nader Bagherzadeh** (Life Fellow, IEEE) received the PhD degree from the University of Texas at Austin, in 1987. He is a professor of computer engineering with the Department of Electrical Engineering and Computer Science, University of California, Irvine, where he served as a chair from 1998 to 2003. He has been involved in research and development in the areas of computer architecture, reconfigurable computing, VLSI chip design, network on chip, 3D chips, sensor networks, computer graphics, memory and embedded systems. Dr. Bagherzadeh has published more than 325 articles in peer-reviewed journals and conferences. His former students have assumed key positions in software and computer systems design companies in the past 30 years.

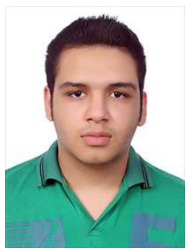

**Ali Sedaghatgoo** received the M.Sc. degree in computer engineering from Amirkabir University of Technology (Tehran Polytechnic), Tehran, Iran, in 2023. He is currently working toward a Ph.D. degree in computer engineering at Sharif University of Technology. His research interests include hardware accelerators, monolithic 3D technology, high-performance computing, and storage systems.

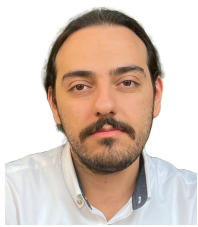

**Amir Mohammad Hajisadeghi** received the M.Sc. degree in computer engineering from Amirkabir University of Technology (Tehran Polytechnic), Tehran, Iran, in 2018. He is currently working toward a Ph.D. degree in computer engineering at Amirkabir University. His research interests include computer architecture, emerging memories, and in-memory computing.